\documentclass[a4paper,twocolumn,english,prl,superscriptaddress]{revtex4}
\usepackage[T1]{fontenc}
\usepackage[latin9]{inputenc}
\usepackage{color}
\usepackage{bm}
\usepackage{amsmath}
\usepackage{amssymb}
\usepackage{graphicx}
\PassOptionsToPackage{normalem}{ulem}
\usepackage{ulem}

\makeatletter

\@ifundefined{textcolor}{}
{%
 \definecolor{BLACK}{gray}{0}
 \definecolor{WHITE}{gray}{1}
 \definecolor{RED}{rgb}{1,0,0}
 \definecolor{GREEN}{rgb}{0,1,0}
 \definecolor{BLUE}{rgb}{0,0,1}
 \definecolor{CYAN}{cmyk}{1,0,0,0}
 \definecolor{MAGENTA}{cmyk}{0,1,0,0}
 \definecolor{YELLOW}{cmyk}{0,0,1,0}
}

\makeatother

\usepackage{babel}
\begin{document}
\title{Chiral excitonics in monolayer semiconductors on patterned dielectric}
\author{Xu-Chen Yang}
\affiliation{Department of Physics, The University of Hong Kong, Hong Kong, China}
\affiliation{HKU-UCAS Joint Institute of Theoretical and Computational Physics at Hong Kong, China}
\author{Hongyi Yu}
\affiliation{Guangdong Provincial Key Laboratory of Quantum Metrology and Sensing and School of Physics and Astronomy, Sun Yat-Sen University (Zhuhai Campus), Zhuhai 519082, China}
\author{Wang Yao}
\email{wangyao@hku.hk}
\affiliation{Department of Physics, The University of Hong Kong, Hong Kong, China}
\affiliation{HKU-UCAS Joint Institute of Theoretical and Computational Physics at Hong Kong, China}
\date{\today}
\begin{abstract}
Monolayer transition metal dichalcogenides feature tightly bound bright excitons at the degenerate valleys, where electron-hole Coulomb exchange interaction strongly couples the valley pseudospin to the momentum of exciton. Placed on periodically structured dielectric substrate, the spatial modulation of the Coulomb interaction leads to the formation of exciton Bloch states with real-space valley pseudospin texture displayed in a mesoscopic supercell. We find this spatial valley texture in the exciton Bloch function is pattern-locked to the propagation direction, enabling nano-optical excitation of directional exciton flow through the valley selection rule. The left-right directionality of the injected exciton current is controlled by the circular polarization of excitation, while the angular directionality is controlled by the excitation location, exhibiting a vortex pattern in a supercell. The phenomenon is reminiscent of the chiral light-matter interaction in nano-photonics structures, with the role of the guided electromagnetic wave now replaced by the valley-orbit coupled exciton Bloch wave in a uniform monolayer, which points to new excitonic devices with non-reciprocal functionalities. 
\end{abstract}
\maketitle

Chiral quantum optical phenomena have recently drawn remarkable interest in a variety of optical systems including photonic crystal waveguides \cite{Sollner15,Le15}, optical fiber \cite{Sayrin15,Petersen14}, whispering-gallery-mode resonators \cite{Shomroni14,Scheucher16}, plasmonic waveguides \cite{Gong18} and metasurfaces \cite{Lin13}. 
As light is tightly confined in these nanophotonic structures, its evanescent longitudinal electric field allows a transverse spin (circular polarization) of light with chirality locked to the propagation direction. The combination of such spin-momentum locking of the guided electromagnetic wave with circularly polarized emitters leads to the chiral light-matter interaction \cite{Lodahl17}. 
The emitter-photon interplay can become non-reciprocal, i.e. forward- and backward-propagating photons interact differently with the emitter of given polarization, and photon emission can even become unidirectional.
The ability to break reciprocity is fundamental for optical devices such as isolators and circulators \cite{Jalas13}, and enables the construction of complex quantum networks as well as the exploration of radically new quantum many-body phenomena mediated by non-reciprocal channels \cite{Lodahl17}.

Valley exciton in monolayer semiconducting transition metal dicalcogenides (TMDs) has been explored as a new type of polarized emitters for the chiral light-matter interface \cite{Gong18,Yang19}. These monolayer semiconductors feature direct band gaps in visible frequency range \cite{Mak10,Splendiani10}, with the conduction and valence band-edges both located at the degenerate $\textbf{K}$ and $-\textbf{K}$ valleys at the Brillouin zone corners. With the strong Coulomb interaction in the 2D geometry, tightly bound Wannier excitons formed in these momentum space valleys dominate the optical responses \cite{Yu15}. Exciton at $\textbf{K}$ ($-\textbf{K}$) valley is interconvertible with a $\sigma+$ ($\sigma-$) polarized photon only \cite{Xiao12,Yao08}. Valley polarized excitons prepared by circular optical pumping can therefore be exploited as chiral emitters for coupling to the spin-orbit coupled electromagnetic wave in nano-photonic structures \cite{Gong18,Yang19}.

The valley exciton also features a pronounced coupling between its centre-of-mass and valley pseudospin degrees of freedom. With the small exciton radius, the exciton dispersion at finite momentum is well split into a transverse (T) and a longitudinal (L) branch, by the sizable electron-hole Coulomb exchange \cite{Yu14,Yu15,Qiu15,Wu15}. The L (T) branch consists of exciton states with in-plane valley pseudospins, i.e. equal superpositions of $\textbf{K}$ and $-\textbf{K}$, that correspond to a linear polarized optical dipole longitudinal (transverse) to exciton momentum \cite{Yu14}. Remarkably, while the T branch has a parabolic dispersion with the regular exciton mass, the L branch is massless with a group velocity proportional to the valley-orbit coupling (Coulomb exchange) strength that is sensitive to the surrounding dielectric \cite{Yu14,Yu15,Qiu15,Wu15}. Such a valley-orbit coupled dispersion implies novel control of exciton transport \cite{Yu14}, as well as the possibility to tailor exciton dispersions through dielectric engineering \cite{Utama19,Forsythe18,Li21,Xu21}.

Here we show a new type of chiral interface where the role of the spin-orbit coupled electromagnetic wave is now played by the valley-orbit coupled exciton Bloch wave in a uniform monolayer placed on structured dielectric substrate. With the spatial modulation in the valley-orbit coupling strength by the periodic dielectric environment, exciton Bloch bands form where the wavefunctions feature spatial texture of valley pseudospin that is pattern-locked to the propagation direction. With the mesoscopic periodicity (O(100) nm) of such texture, nano-optical excitation in selected energy window can directly inject directional exciton flow in the 2D plane. The left-right directionality of exciton current is controlled by the circular polarization of excitation, while the angular directionality is controlled by the excitation location. The current injection rate as a function of the excitation location in a supercell exhibits a vortex (anti-vortex) pattern for left- (right-) handed circular polarization. 

The valley excitons in monolayer TMDs are described by the Hamiltonian $\hat{H}=\sum_{\textbf{k},\tau}\frac{\hbar^2k^2}{2M_0} \hat{B}_{\textbf{k},\tau}^\dagger\hat{B}_{\textbf{k},\tau}+\hat{H}_\text{ex}$, where $\hat{B}^\dagger_{\textbf{k},\tau}$ creates an exciton with  center-of-mass (COM) momentum $\textbf{k}=(k\cos\theta_\textbf{k},k\sin\theta_\textbf{k})$ and  valley index $\tau=+,-$. The electron-hole exchange $\hat{H}_\text{ex}$ can be generally expressed in terms of the exciton operators \cite{Yu14}: $\hat{H}_\text{ex} = \sum_{\textbf{k},\textbf{k}^\prime,\tau,\tau^\prime} \hat{B}^\dagger_{\textbf{k},\tau}\hat{B}^{}_{\textbf{k}^\prime,\tau^\prime} (\textbf{k}\cdot\textbf{d}_{\tau})(\textbf{k}^\prime\cdot\textbf{d}^\ast_{\tau^\prime})V(\textbf{k}-\textbf{k}^\prime,\frac{m_\text{h}}{M_0}\textbf{k}+\frac{m_\text{e}}{M_0}\textbf{k}^\prime)$, where $\textbf{d}_{\tau}$ is the optical dipole of the exciton and $V(\textbf{Q},\textbf{q})\equiv\int\text{d}\textbf{r}_\text{eh}\text{d}\textbf{R}\text{e}^{-\text{i}\textbf{q}\cdot\textbf{r}_\text{eh}}\text{e}^{-\text{i}\textbf{Q}\cdot\textbf{R}}V(\textbf{R},\textbf{r}_\text{eh})$ is the Fourier transform of the electron-hole Coulomb interaction. In the patterned dielectric surrounding, $V$ depends on both the exciton COM coordinate $\textbf{R}\equiv\frac{m_\text{e}}{M_0}\textbf{r}_\text{e}+\frac{m_\text{h}}{M_0}\textbf{r}_\text{h}$, and the electron-hole relative coordinate $\textbf{r}_\text{eh}\equiv\textbf{r}_\text{e}-\textbf{r}_\text{h}$~\cite{Chernikov14,Raja19}.
We have dropped the short-ranged part of the exchange interaction, which can be regarded as a constant for small value of $k$ and $k^\prime$ \cite{Yu14}. 

It is convenient to switch to the basis of L and T branch excitons, defined as $\hat{B}^\dagger_{\textbf{k},\text{L}}\equiv\frac{1}{\sqrt{2}}\left(\text{e}^{-\text{i}\theta_\textbf{k}}\hat{B}^\dagger_{\textbf{k},+}+\text{e}^{\text{i}\theta_\textbf{k}}\hat{B}^\dagger_{\textbf{k},-}\right)$ and $\hat{B}^\dagger_{\textbf{k},\text{T}}\equiv\frac{1}{\sqrt{2}}\left(\text{e}^{-\text{i}\theta_\textbf{k}}\hat{B}^\dagger_{\textbf{k},+}-\text{e}^{\text{i}\theta_\textbf{k}}\hat{B}^\dagger_{\textbf{k},-}\right)$.
Noting that the exciton has a valley contrasted circularly polarized dipole\cite{Xiao12}, $\textbf{d}_{\tau}=\frac{D}{\sqrt{2}}(\text{i}\tau,1)$, the exchange term reduces to
\begin{equation}
\hat{H}_\text{ex}=D^2\sum_{\textbf{k},\textbf{k}^\prime}kk^\prime V\left(\textbf{k}-\textbf{k}^\prime,\frac{m_\text{h}}{M_0}\textbf{k}+\frac{m_\text{e}}{M_0}\textbf{k}^\prime\right)\hat{B}^\dagger_{\textbf{k},\text{L}}\hat{B}^{}_{\textbf{k}^\prime,\text{L}}
\label{eq:Ham}
\end{equation}
The Coulomb exchange interaction only affects the L branch exciton that emit photon of linear polarization longitudinal to the momentum $\textbf{k}$, while the T branch is unaffected. Notably, this conclusion is independent on the form of $V(\textbf{R},\textbf{r}_\text{eh})$.
On homogeneous substrate, the Coulomb interaction depends on $\textbf{r}_\text{eh}$ only, so $V(\textbf{Q},\textbf{q})=\delta_{\textbf{Q},0} \frac{2\pi e^2}{\varepsilon q}$. 
The exciton Hamiltonian becomes $\hat{H}=\sum_\textbf{k} \frac{\hbar^2k^2}{2M_0} \hat{B}_{\textbf{k},\text{T}}^\dagger\hat{B}_{\textbf{k},\text{T}}+  \left( \frac{\hbar^2k^2}{2M_0} + \frac{2\pi e^2}{\varepsilon} D^2 k  \right) \hat{B}_{\textbf{k},\text{L}}^\dagger\hat{B}_{\textbf{k},\text{L}}$, i.e., the T branch has the regular dispersion determined by exciton mass $M_0$, while the L branch becomes massless with a group velocity determined by the effective dielectric constant $\varepsilon$ (c.f. Fig. 1(a)). 

\begin{figure}
\includegraphics[scale=0.38]{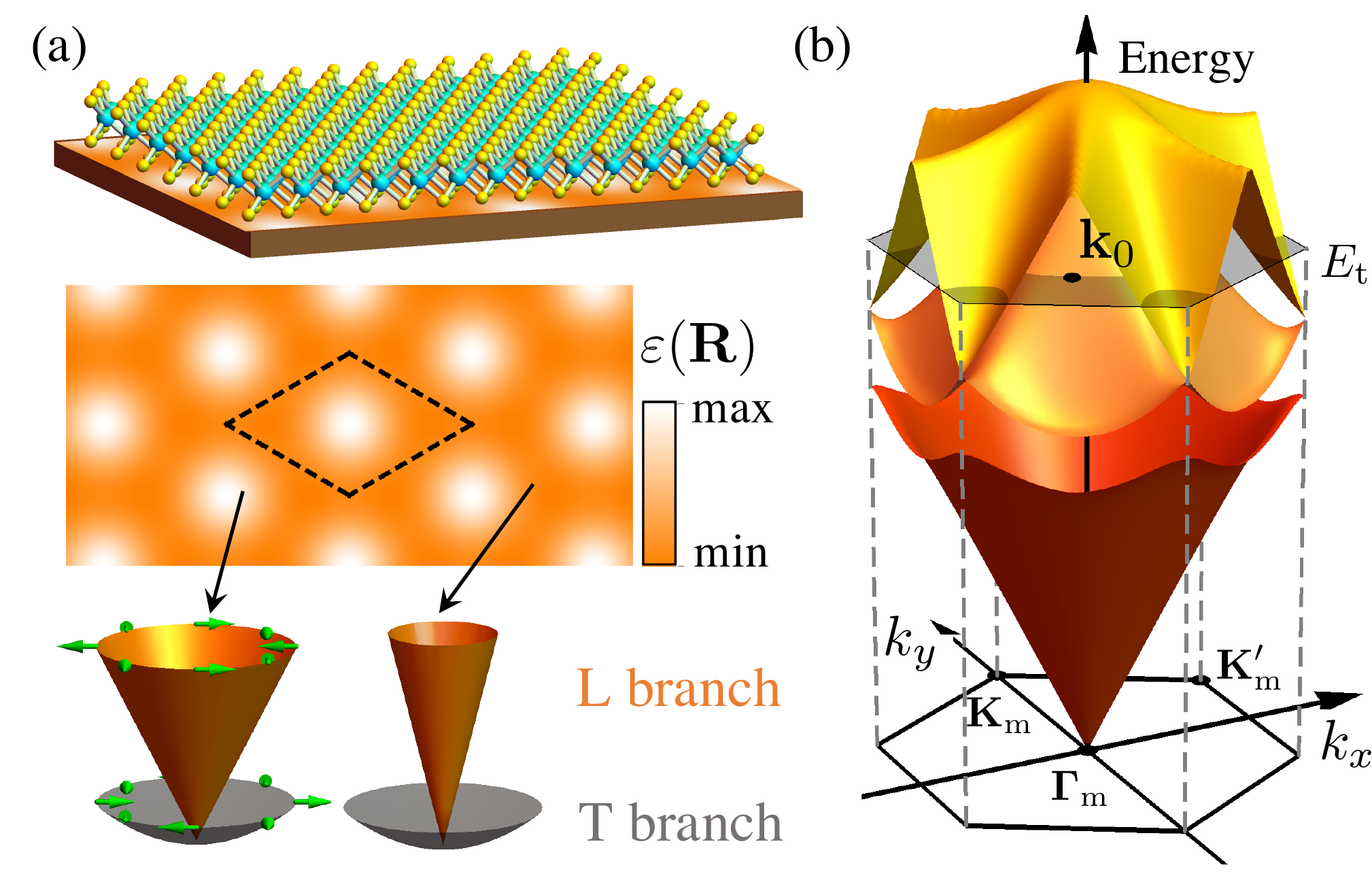}
\caption{(a) Schematic of a monolayer TMDs on patterned substrate where the dielectric $\varepsilon$ has a triangular superlattice profile as a function of location $\textbf{R}$. The bottom plots show local dispersions of exciton at the high and low $\varepsilon$ regions. Because of the sizable electron-hole Coulomb exchange, exciton dispersions split into L (orange) and T (grey) branches featuring opposite momentum-space textures of valley pseudospin (green arrows). The L branch has a massless dispersion with velocity $\propto \varepsilon^{-1}$, while the T branch has the usual exciton mass independent of $\varepsilon$. For the L branch exciton, the dependence on $\varepsilon$, together with the modulation $\varepsilon(\textbf{R})$, leads to formation of mini-bands. (b) Typical example of L branch dispersion shown in the superlattice mini-zone (the same as Fig. 3(a) where parameters used in the calculation are listed).
\label{fig:dielectric_superlattice}}
\end{figure}

Now we consider placing the monolayer on a patterned substrate (Figure \ref{fig:dielectric_superlattice}(a)), where $\varepsilon$ gets spatially modulated with a periodicity $l \sim O(100)$nm. The local dispersion differs from place to place for the L branch exciton, which leads to formation of mini-bands. In the superlattice defined by the patterned dielectric, the Coulomb interaction is of the general form $V(\textbf{Q},\textbf{q})=\sum_\textbf{g}\delta_{\textbf{g},\textbf{Q}}V_{\textbf{g}}(\textbf{q})$, expanded in terms of the reciprocal lattice vectors $\textbf{g}$ (c.f. supplementary information). For smoothly varying $\varepsilon$, keeping the lowest several $\textbf{g}$ can be sufficient. Eq. (1) now consists of $\hat{B}_{\textbf{k+g},\text{L}}^\dagger\hat{B}_{\textbf{k},\text{L}}$, i.e. the exciton momentum scattering by the dielectric superlattice, responsible for the mini-gap opening of the L branch (Fig.\ref{fig:dielectric_superlattice}(b)). 

\begin{figure*}
\includegraphics[scale=0.4]{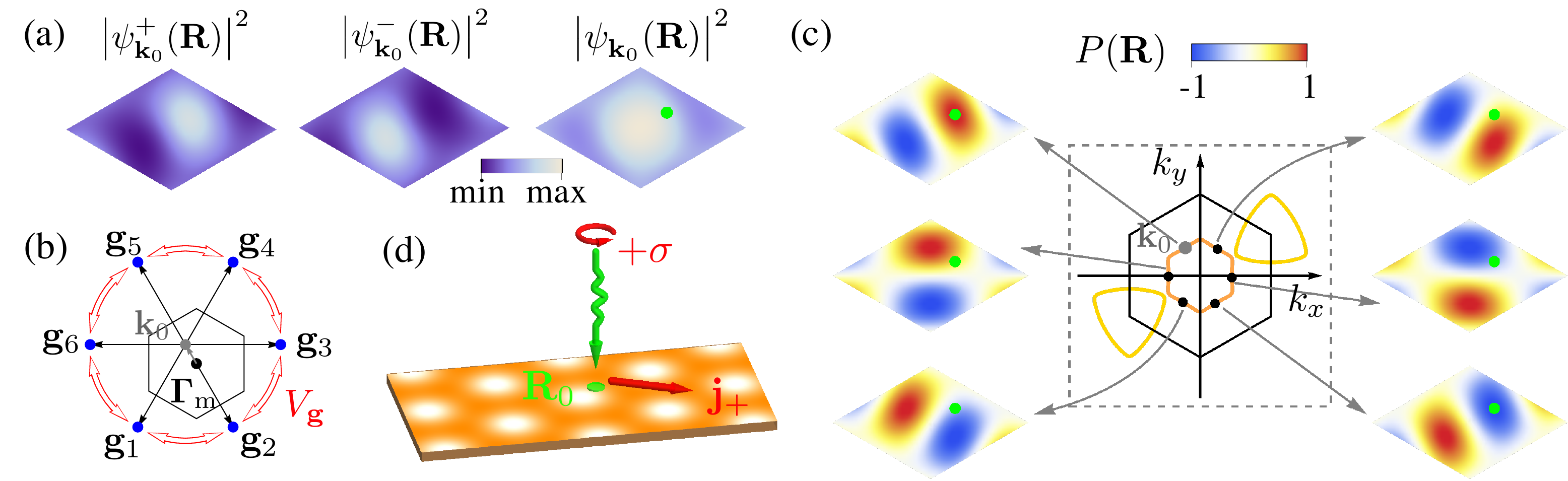}
\caption{(a) Example of L exciton Bloch function (at wavevector $\textbf{k}_0$ denoted by the dot in Fig. 1(b)). 
Left to right: probability density of the $\textbf{K}$, $-\textbf{K}$ valley components, and their sum, shown in a supercell (c.f. dashed diamond in Fig. 1(a)).  (b) This Bloch function is predominantly formed by the linear combination of the six modes at $\textbf{k}_0 + \textbf{g}_i$ of the unfolded L branch, which have nearly degenerate kinetic energy, now hybridized by Fourier component $V_{\textbf{g}}$ of the electron-hole Coulomb exchange. $\textbf{g}_i$ are reciprocal superlattice vectors. (c) Spatially resolved valley polarization $P_\textbf{k}(\textbf{R})$ in the supercell, shown for six exciton Bloch functions on the contour of energy $E_\text{t}$ (c.f. Fig. 1(b)). The orientation of the texture is locked to the superlattice crystal momentum. (d) Through circular polarized nano-optical excitation at $\textbf{R}_0$, out of the six Bloch states shown in (c), the one at  $\textbf{k}_0 $ is selectively excited because of its valley pseudospin texture, leading to a directional exciton follow.
\label{fig:wavefunction}}
\end{figure*}

The unfolded L branch features a valley pseudospin texture in the momentum space (Fig.1(a)). Remarkably, the band folding combined with a momentum-space pseudospin texture in general leads to superlattice Bloch function with a real-space pseudospin texture in a supercell. Consider, for example, the Bloch function at an arbitrary crystal momentum $\textbf{k}_0$ in the second lowest mini-band (c.f. Fig. 1(b)), which is a linear superposition of the six L exciton modes $\hat{B}_{\textbf{k}_0 + \textbf{g}_i,\text{L}}$ with nearly degenerate kinetic energy hybridized by $\hat{H}_{\text{ex}}$ (c.f. Fig. \ref{fig:wavefunction}(b)). 
Using $\psi^\tau_{\textbf{k}_0}(\textbf{R})$ to denote the valley $\tau$ component of this Bloch function, the perturbation expansion leads to (c.f. supplementary information),
\begin{eqnarray}
|\psi^\tau_{\textbf{k}_0}(\textbf{R})|^2 &=& \frac{1}{2}-\frac{g^2 \eta}{\textbf{k}_0\cdot\textbf{g}_{i}}\cos(\textbf{g}_{i}\cdot\textbf{R}+\tau\frac{\pi}{3}) \notag \\ 
&& -\frac{g^2 \eta}{\textbf{k}_0\cdot\textbf{g}_{j}}\cos(\textbf{g}_{j}\cdot\textbf{R}-\tau\frac{\pi}{3}), 
\end{eqnarray}
where $\textbf{g}_{i}$, $\textbf{g}_{j}$ are two reciprocal vectors next-nearest to $\textbf{k}_0$ (i.e. $\textbf{g}_{4,6}$ in Fig.~2(b)), and $\eta$ is a dimensionless parameter. Notably, the two valley components $\psi^+_{\textbf{k}_0}(\textbf{R})$ and $\psi^-_{\textbf{k}_0}(\textbf{R})$ now display distinct probability distribution in the supercell, pattern-locked to the momentum direction. Fig. 2(a) plots the numerically calculated Bloch function for a typical dielectric superlattice, in excellent agreement with Eq. (2). 
We use $P_\textbf{k}(\textbf{R})=\frac{|\psi_{\textbf{k}}^+(\textbf{R})|^2-|\psi_{\textbf{k}}^-(\textbf{R})|^2}{|\psi_{\textbf{k}}^+(\textbf{R})|^2+|\psi_{\textbf{k}}^-(\textbf{R})|^2}$ to quantify the spatial resolved valley polarization. 
As shown in Fig. \ref{fig:wavefunction}(c), the spatial pattern of the valley polarization $P_\textbf{k}(\textbf{R})$ for the exciton Bloch states rotates with the direction of $\textbf{k}$. 

This makes possible a chiral interface to inject directional exciton flow in the 2D plane through a circularly polarized nano-optical excitation (Fig. \ref{fig:wavefunction}(d)). Consider an excitation field profile enhanced by a chiral nanostructure~\cite{Qiu18, Gao21}, 
which can have a spatial extension of $O(10)$ nm, small compared to the superlattice period $l$, but orders larger than the monolayer lattice constant ($\sim 0.3$ nm). By the valley optical selection rule, the $\sigma+$ polarized optical excitation couples only to the $\tau=+$ valley component of the exciton Bloch state, so the excitation rate is determined by the location-dependent probability density $|\psi^+_{\textbf{k}}(\textbf{R})|^2$. At given excitation energy, while all $\textbf{k}$ points on the energy contour can be excited in general, the excitation rate can have a strong dependence on the direction of $\textbf{k}$, with the angular distribution controlled by the excitation location. 
For example, among the six Bloch states on the energy contour shown in Fig. \ref{fig:wavefunction}(c), $\sigma+$ polarized excitation at the green dot will preferentially populate the one at $\textbf{k}_0$, so the injected excitons correspond to a net flow along the red arrow in Fig. 2(d). Likewise, the $\sigma-$ polarized excitation at the same location inject an opposite exciton flow. The angular directionality is further controlled by the excitation location in the supercell, evidenced by the $\textbf{R}$ dependence in Eq. (2) (see also Fig. 2(c)). 

\begin{figure*}
\includegraphics[scale=0.45]{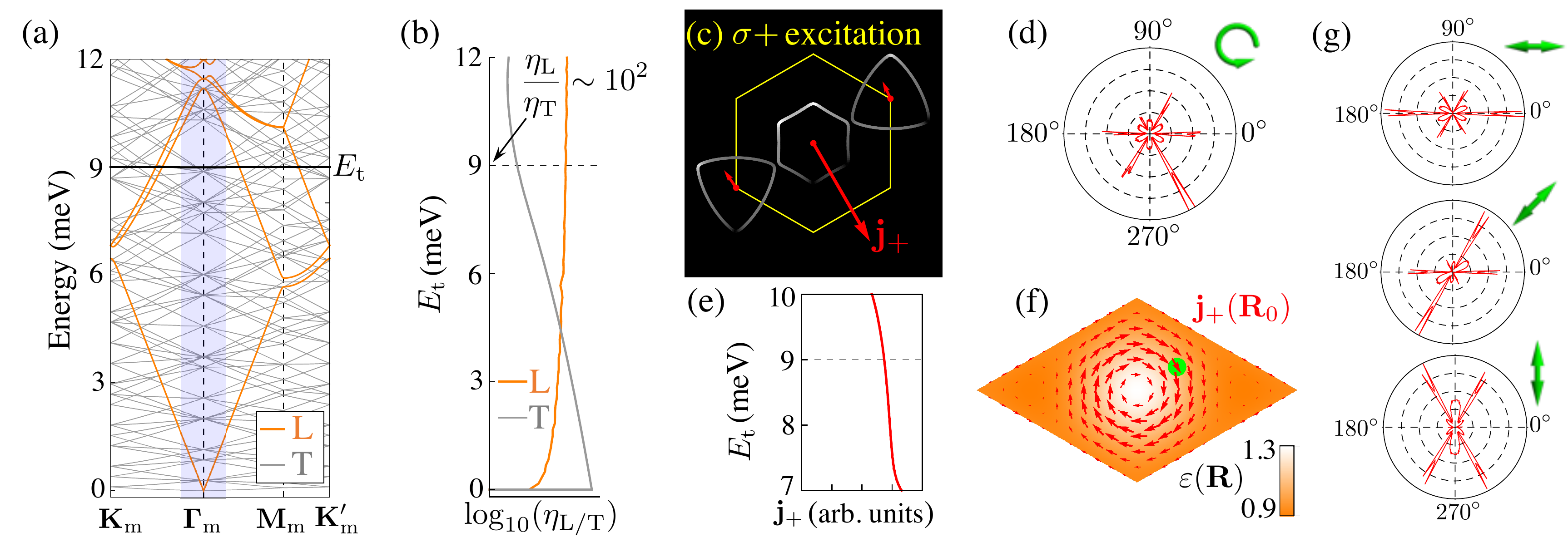}
\caption{(a) Exciton dispersion in dielectric superlattice characterized by $\varepsilon^{-1}(\textbf{R}) = \varepsilon_c^{-1}\left[ \alpha(0) +  \sum_{i=1}^6 \alpha(\textbf{g}_i)  \exp (i \textbf{g}_i \cdot\textbf{R})\right]$, colour coded by the projection of the Bloch functions to the T (grey) and L branch (orange) modes. $\alpha(0)=1$, $\alpha(\textbf{g}_i)=-0.04$, corresponding to a dielectric modulation between $0.9\varepsilon_c$ and $1.3\varepsilon_c$. The superlattice period $l=100$ nm, and electron-hole exchange strength $J=1$ eV (c.f. Eq. (3)). (b-f) Injection of excitons by a $\sigma+$ polarized nano-optical excitation with a field profile $\mathcal{G}(\textbf{R})=\text{exp}\left[-2(\textbf{R}-\textbf{R}_0)^2/ w^2\right]$, with $w=10$ nm centred at $\textbf{R}_0$. (b) $\eta_{\text{L}/\text{T}}$ denotes the overall injection rate of L/T excitons as a function of excitation energy $E_\text{t}$. (c) Population distribution of the injected L excitons in momentum space at the excitation energy $E_\text{t} = 9$ meV (denoted by horizontal line in (a) and (b)). The anisotropic distribution in momentum space corresponds to an exciton current, primarily contributed by the inner contour (c.f. red arrows). (d) Angular distribution of the injected exciton flow. (e) The current injection rate $\textbf{j}_+$ as a function of $E_\text{t}$. (f) $\textbf{j}_+$ as a function of excitation location $\textbf{R}_0$ in a supercell, exhibiting a vortex pattern. Green dot marks the excitation location for the plots of (b-e). (g) Angular distributions of the injected exciton flow under various linearly polarized excitation at $\textbf{R}_0$. 
\label{fig:numerical}}
\end{figure*}

\begin{figure}
\includegraphics[scale=0.39]{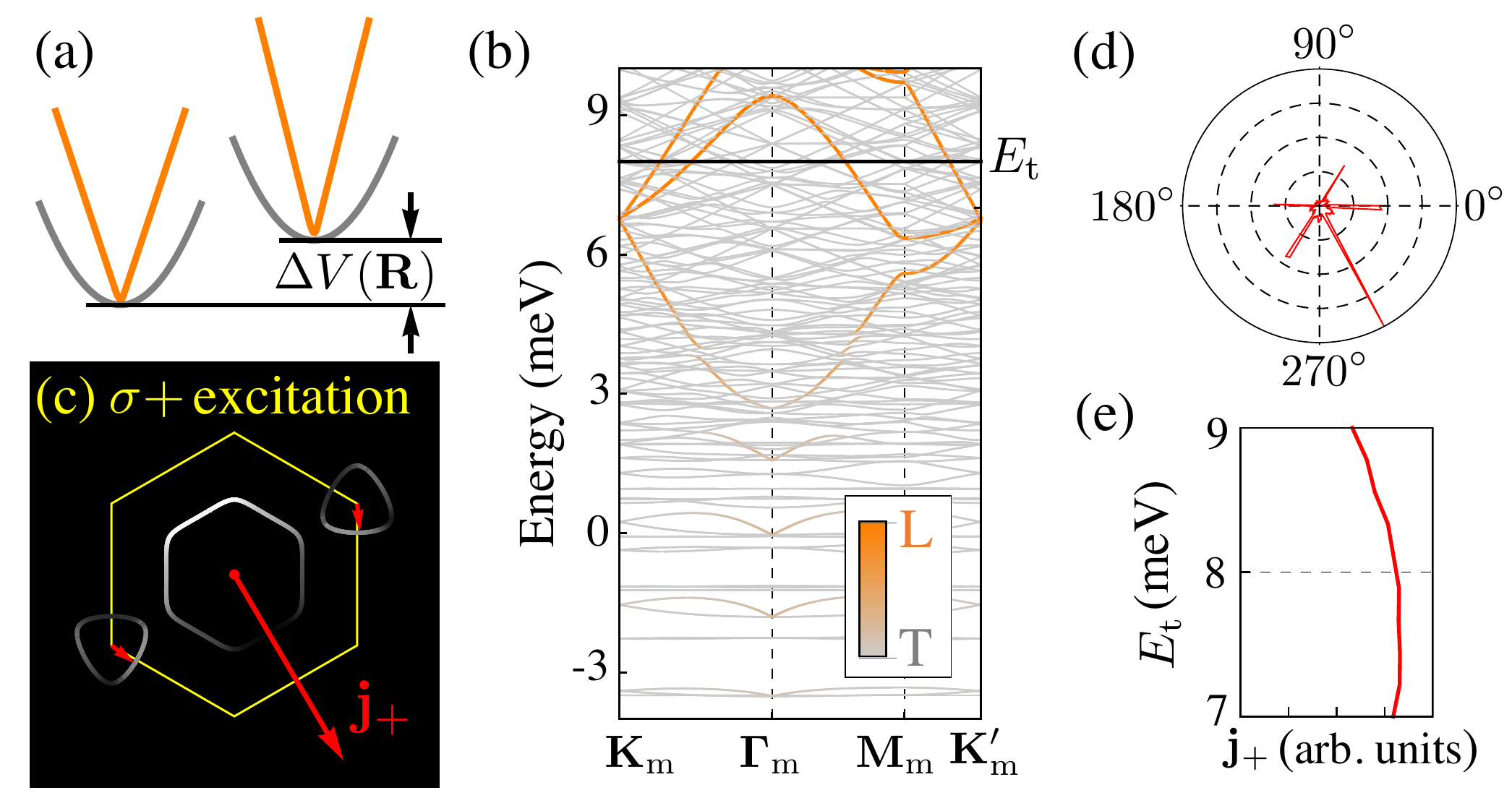}
\caption{(a) The patterned dielectric can also lead to a weak spatial modulation of exciton energy $\Delta V(\textbf{R})$ independent of valley pseudospin. (b) Exciton mini-bands taking into account $\Delta V(\textbf{R})$ with a modulation amplitude of 9 meV. Other parameters are the same as in Fig. 3. The proportion of the Bloch states on the L and T modes are colour-coded. Above 6 meV, there are mini-bands predominantly from the L branch (orange). (c) Population distribution of the injected L excitons in momentum space at $E_t = 8$ meV (c.f. (a)). (d) Angular distribution of the injected exciton flow. (e) The current injection rate $\textbf{j}_+$ as a function of excitation energy $E_\text{t}$.  
\label{fig:modulation}}
\end{figure}

For a quantitative analysis of the chiral interface function, we adopt a screened form of the Coulomb interaction $V_{\textbf{g}}(\textbf{q})=\alpha(\textbf{g})\frac{2\pi\text{e}^2}{\varepsilon_c}\frac{1}{ \lambda g+q}$. $\lambda g$ is the screening wavevector (c.f. supplementary information), and $\alpha(\textbf{g})$ is the Fourier coefficient of the spatially modulated dielectric $\varepsilon^{-1}(\textbf{R}) = \varepsilon_c^{-1}\left[ \alpha(0) +  \sum_{i=1}^6 \alpha(\textbf{g}_i)  \exp (i \textbf{g}_i \cdot\textbf{R})\right]$, the six $\textbf{g}_i$ shown in Fig. 2(b). Eq. (1) then becomes, 
\begin{equation}
\hat{H}_\text{ex}=\frac{J}{K}\sum_{\textbf{k},\textbf{k}^\prime,\textbf{g}}\delta_{\textbf{k}^\prime,\textbf{k}+\textbf{g}} \frac{kk^\prime \alpha(\textbf{g})}{\lambda g +\left|\frac{m_\text{h}}{M_0}\textbf{k}+\frac{m_\text{e}}{M_0}\textbf{k}^\prime\right|}\hat{B}^\dagger_{\textbf{k},\text{L}}\hat{B}^{}_{\textbf{k}^\prime,\text{L}}
\label{eq:Ham1}
\end{equation}
where $J\equiv\frac{2\pi\text{e}^2D^2}{\varepsilon_c}K\sim1$ eV can be extracted from first principle wavefunctions and exciton spectrum~\cite{Yu14,Qiu15,Wu15}. We have set the dielectric modulation in a range between $0.9\varepsilon_c$ and $1.3\varepsilon_c$, corresponding to $\alpha(0)=1$, $\alpha(\textbf{g}_i)=-0.04$, where the variation in exciton optical resonance is neglected. 

Fig.\ref{fig:numerical} (a) shows an example of the exciton mini-bands calculated for the superlattice period $l=100$ nm, and $\lambda=1$. Each eigenstate is colour coded by the projection of the Bloch function to the T branch (grey) and L branch (orange) modes. As expected, only the L branch is affected by the superlattice modulation of the dielectric through the electron-hole exchange interaction $\hat{H}_\text{ex}$. The spectrum is not sensitive to the parameter $\lambda$, while the energy scales inversely with the superlattice period $l$. Examples of the Bloch functions in the second lowest mini-bands at the energy $E_\text{t}$ have been shown in Fig. 2(a) and 2(c), which are well described by Eq. (2). Except for the lowest mini-band, the pattern-locking of the real-space valley texture to the direction of superlattice crystal momentum $\textbf{k}$ is generally found. This suggests a large energy window to explore the chiral interface function. 

The excitation rate of an exciton Bloch state is given by $\zeta(\textbf{k})=\left|\left\langle\psi_{\textbf{k}}^+|\mathcal{G}\right\rangle\right|^2$, where $\mathcal{G}$ characterises the spatial profile of a $\sigma +$ polarized nano-optical excitation. The main results below are insensitive to the detail of $\mathcal{G}$ as long as its spatial extension is small compared to $l$. For simplicity, we use a Gaussian $\mathcal{G}(\textbf{R})=\text{exp}\left[-2(\textbf{R}-\textbf{R}_0)^2/w^2\right]$, with $w=10$ nm. 
Fig. \ref{fig:numerical}(b) compares the overall injection rate of L and T branch excitons as function of excitation energy, $\eta(E_\text{t}) = \gamma \int_{E_\text{t}}\frac{\zeta(\textbf{k})}{|\nabla_\textbf{k}E|}\text{d}k$,
where $\gamma$ is proportional to the field intensity. While the excitation rate of the T branch is higher at the spectral edge because of the larger density of state, starting from the energy window of the second L branch mini-band, the excitation of the L branch dominates, as the T branch modes correspond to large COM momentum that results in diminishing $\zeta(\textbf{k})$. 

Fig. \ref{fig:numerical}(c) presents the momentum space distribution of the injected L excitons at excitation energy $E_\text{t}=9$ meV, where the excitation center $\textbf{R}_0$ is denoted by the green dot in Fig. \ref{fig:numerical}(f). As expected, a highly anisotropic distribution is injected by the nano-optical excitation, which corresponds to a directional exciton flow. Fig. \ref{fig:numerical}(d) plots the angular distribution of the exciton flow. The injection rate of the overall exciton current is given by,
\begin{equation}
    \textbf{j}_+=\gamma \int_{E_\text{t}}\zeta(\textbf{k}) \frac{\nabla_\textbf{k}E}{|\nabla_\textbf{k}E|}\text{d}k,
\label{eq:current}
\end{equation}
For the excitation considered in Fig. 3(c), $\textbf{j}_+$ is predominantly contributed by the inner energy contour. Fig. \ref{fig:numerical}(f) plots the current injection rate as a function of the excitation location $\textbf{R}_0$ in a supercell, which exhibits a vortex pattern. In Fig. \ref{fig:numerical}(g), we also examined the angular directionality upon linearly polarized excitation, where the bidirectional polar distribution is also controllable by the polarization angle. It is worth noting that the shape of the inner energy contour (Fig. \ref{fig:numerical}(c)) has resulted in collimation of the exciton beam along 6 directions.

The quasiparticle bandgap and exciton binding energy are also sensitive to $\varepsilon$~\cite{Ugeda14,Raja17,Stier16}. 
For 1s exciton, nevertheless, various experiments have shown that the optical resonance (dispersion edge) is largely unaffected by the surrounding dielectric \cite{Stier16,Raja19,Xu20a,Lin14}
as the bandgap and binding energy dependence on $\varepsilon$ predominantly cancel each other~\cite{Qiu13,Cho18,Gao16}.
Experiments in Ref.~\cite{Raja19}
show that upon the dielectric variation that has resulted in $\sim 30 - 40 \%$ change in exciton binding energy, the change in 1s exciton optical resonance is only a few meV. This effect can be accounted by adding a superlattice potential $\Delta V(\textbf{R})$ independent of the exciton valley pseudospin (c.f. Fig. 4(a)). 
Fig. \ref{fig:modulation}(b) plots the exciton mini-bands, with $\Delta V(\textbf{R})$ having a modulation amplitude of 9 meV. This  potential modulation leads to folding of the T branch, and some hybridization of the L and T branches in the low energy sector only. In the spectral window above 6 meV, one can clearly identify mini-bands arising predominantly from the L branch (orange), where the dispersion largely resembles the ones in the absence of $\Delta V(\textbf{R})$ (c.f. Fig. 3(a)). Fig. \ref{fig:modulation}(c) plots the momentum space distribution of the excitons injected at excitation energy $E_\text{t}=8$ meV, and Fig. \ref{fig:modulation}(d) shows the angular distribution of the exciton flow. Comparing with the plots of Fig. 3(c) and (d), one finds the directionality and the order of magnitude of the exciton current inject remains largely unaffected by the presence of $\Delta V(\textbf{R})$.

In summary, we show the strong valley-orbit coupling combined with the superlattice modulation leads to periodic spatial texture of valley pseudospin correlated with wavevector in exciton Bloch functions. With the mesoscopic superlattice periodicity resolvable in nano-optical excitation, such spatial valley texture can be exploited to realize a chiral interface for injection of exciton flow. The superlattice modulation naturally arises when the monolayer is placed on a patterned dielectric substrate~\cite{Forsythe18, Li21},  
including a typical photonic crystal slab, where the excitonic and photonic properties can be simultaneously engineered. A novel alternative for the superlattice modulation of the dielectric surrounding of monolayer semiconductor is also demonstrated recently using graphene/hBN moire superlattice~\cite{Xu21}. These point to exciting opportunities towards excitonic devices with non-reciprocal functionalities. 

The work is mainly support by the University Grants Committee/Research Grant Council of the Hong Kong SAR (AoE/P-701/20), the HKU Seed Funding for Strategic Interdisciplinary Research, and the Croucher Senior Research Fellowship. H.Y. acknowledges support by the Department of Science and Technology of Guangdong Province in China (2019QN01X061).

%

\end{document}